\documentclass[aps,prl,twocolumn]{revtex4-1}
\usepackage{amsmath}
\usepackage{amssymb}
\usepackage{graphics}
\usepackage{epsfig}
\usepackage{color}
\usepackage{graphicx}
\usepackage{color}
\begin{document}
\title{Noisy  NF-$\kappa$B oscillations stabilize and sensitize cytokine signaling in space.}

%\title{Noisy  NF-$\kappa$B oscillations stabilize cytokine signaling.}

%\title{The role of noisy  NF-$\kappa$B oscillations in spatio-temporal cytokine patterns.} 

\author{Sirin W. Gangstad$^{1,2}$}                                	% Forfatter p? artiklen (brug en ny \author{} kommando for hver forfatter)
\author{Cilie W. Feldager$^{1,2}$}
\author{Jeppe Juul$^1$}                                    %  ...
\author{Ala Trusina$^1$}                                    	%  ...
\affiliation{$^1$Niels Bohr Institute, University of Copenhagen}
\affiliation{$^2$These authors contributed equally to this work.}   % Forfatternes tilh?rsforhold

\begin{abstract}  
NF-$\kappa$B is a major transcription factor mediating inflammatory response. In response to pro-inflammatory stimulus, it exhibits characteristic response -- a pulse followed by noisy oscillations in concentrations of considerably smaller amplitude.
NF-$\kappa$B is an important mediator of cellular communication, as it is both activated by and upregulates production of cytokines, signals used by white blood cells to find the source of inflammation. While the oscillatory dynamics of NF-$\kappa$B has been extensively investigated both experimentally and theoretically, the role of the noise and the lower secondary amplitude has not been addressed. 

We use a cellular automaton model to address these issues in the context of spatially distributed communicating cells.
We find that noisy secondary oscillations stabilize concentric wave patterns, thus improving signal quality. Furthermore, 
both lower secondary amplitude as well as noise in the oscillation period might be working against chronic inflammation, the 
state of self-sustained and stimulus-independent excitations.   

Our findings suggest that the characteristic irregular secondary oscillations of lower amplitude are not accidental. On the contrary, they might have evolved to increase robustness of the inflammatory response and the system's ability to return to pre-stimulated state.
\end{abstract}

\maketitle                                                	% Laver overskrift mm. for artiklen

\section{Introduction}
The regulatory network of NF-$\kappa$B is an important constituent of the immune system as it regulates hundreds of genes in response to extracellular stimuli such as pathogens, cytokines and stress \cite{Brasier2006, Hoffmann2006}. These responses include apoptosis, cell proliferation, and inflammatory response \cite{Pahl1999, Brasier2006}.

During inflammatory response, the NF-$\kappa$B network is activated by an increase in extracellular concentration of cytokines; small signaling molecules commonly used in intercellular communication. Once activated, NF-$\kappa$B up-regulates the cells' own cytokine production, thereby amplifying the external signal and passing it to the neighbouring cells \cite{Lee2009, Kasper2010} either by diffusion or through gap-junctions -- channels formed by physically interacting cells \cite{Kasper2010}. 

When passing the signal from one cell to the next in this manner, the tissue acts as an excitable medium. Recent theoretical research has shown that propagating elevated cytokine concentrations as waves through this ``excitable tissue'' might represent an optimal way of passing the signal to the blood vessels, where the cytokines are absorbed \cite{Yde2011a,Yde2011c}. When neutrophils (white blood cells) detect the cytokines in the blood stream, they will follow the cytokine concentration gradient back to the site of infection \cite{Geiger2003}, in order to destroy or contain the cause of the infection \cite{Witko-Sarsat2000}.

Single cell measurements \cite{Tay2010, Lee2009} and modeling approaches \cite{Mengel2012, Lipniacki2004} revealed that the concentration of active NF-$\kappa$B oscillates with an initial high amplitude peak and several consecutive lower amplitude peaks. The experiments also revealed a considerable amount of noise in this response, and simulations suggest that the noise is in fact induced by an inherent component of the network \cite{Ashall2009, Marcello2010, Turner2010, Lipniacki2007}.
The effect of noise on the wave propagation of cytokines through the tissue has not yet been elucidated, but the presence of a noise-inducing component suggests that the noise itself may play a key role in the immune response. This has motivated us to ask the following questions:
How does noise affect wave propagation? 
Is the system equally sensitive to irregularities in the period of oscillations as to irregularities in the refractory period? 
Does noise contribute to the onset of chronic inflammation -- the state where waves are self-sustained and do not depend on stimulus?

\section{the Model}
\begin{figure}[t!]
\includegraphics[width=0.5\textwidth]{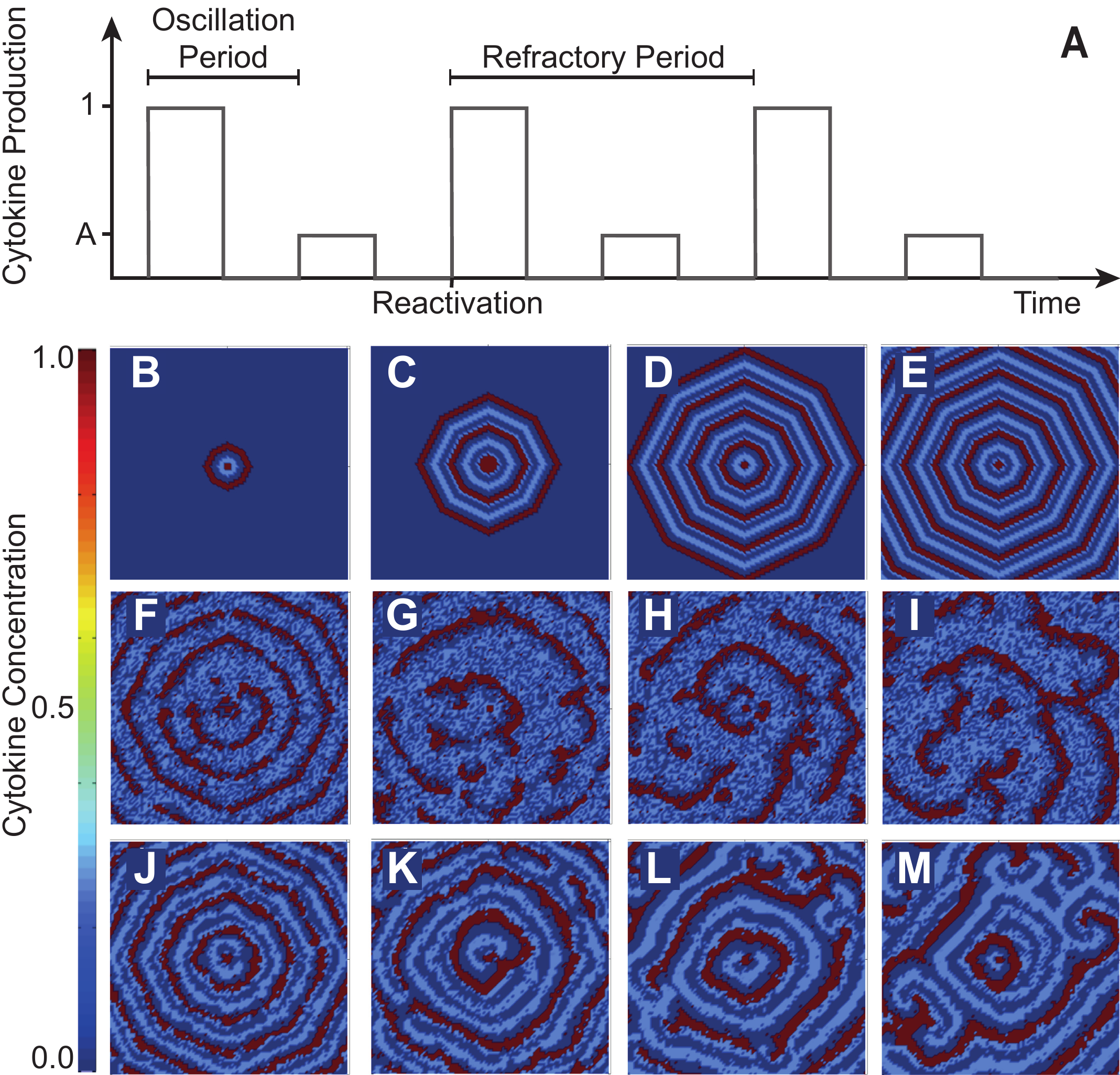}
\caption{ \textbf{A}: The oscillatory cytokine production of a single cell subject to constant stimulation. The amplitude of the initial peak is normalized to 1 while all secondary peaks have amplitude $A$. When the cell leaves refractory period, it is here immediately re-excited to its initial peak.
 \textbf{B-E}: show wave propagation in an ideal system without any noise. 
\textbf{F-I}: have $\eta_o =40$ \% noise in the oscillation period and no noise in the refractory period, $\eta_r =0$. 
\textbf{J-M}: show the system with no noise in the oscillation period, $\eta_o =0$, but $\eta_r =10$ \% noise in the refractory period. 
The secondary amplitude is $A=0.25$ in all panels A-M. In the time series F-I and J-M, snapshots are taken at t = 20, 60, 100 and 200 hours. We see that both kinds of noise gradually generate spirals of active cells, which may mislead neutrophils and cause chronic inflammation. 
}
\label{fig:TimeSeries}
\end{figure}

The tissue is modeled as an excitable media using a parallel cellular automaton, an algorithm frequently used as a mathematical idealization of biological self-organizing systems \cite{Wolfram1983, Bub2002, Bub2005, Sawai2005, Marr2006}. Our model comprises $101 \times 101$ cells placed on a square grid. All cells are initially inactive, but once activated by their neighbouring cells they start producing cytokines in an irregularly oscillating fashion with an initial high peak followed by several lower peaks (see Fig. \ref{fig:TimeSeries}a). The amplitude of the initial peak is normalized to 1 and the mean oscillation period is estimated from current literature to be 100 minutes \cite{Ashall2009}, corresponding to 12 time steps in the model. The amplitude $A$ of all secondary peaks and the relative size $\eta_o$ of the gaussian noise in the oscillation period are kept as free parameters, but retained similar to values experimentally measured in nuclear NF-$\kappa$B. 

A cell is activated if the combined cytokine production of its 8 nearest neighbours, the Moore-neighbourhood, exceeds a threshold $\theta=3$. That is, a cell can be activated if 3 of its neighbours are in their first peak of cytokine production, or if e.g. 2 neighbours are in their first peak and $1/A$ other neighbours are in a secondary peak. When a cell is activated, it enters a refractory period in which it is completely unresponsive to the local cytokine concentration, and will produce cytokines at its own pace. The mean refractory period is estimated from current literature to be 200 minutes \cite{Nelson2004a}, but here is also added a gaussian noise of relative size $\eta_r$, kept as a free parameter. Once out of the refractory period, the cell becomes susceptible to input signals again, and if the local cytokine concentration should rise above the threshold, the cell will re-initiate the cytokine production, responding with the initial high peak followed by several low peaks (see Fig. \ref{fig:TimeSeries}a).

A stimulus of $3 \times 3$ constantly excited cells in the middle of the grid was added to act as a pathogen and initiate a train of cytokine waves propagating through the tissue.

\section{Results}
An example of the effect of noise in the wave propagation is shown in Fig. \ref{fig:TimeSeries}. Here, noise is added to the period of oscillations (Fig. \ref{fig:TimeSeries}F-I) and to the duration of the refractory period (Fig. \ref{fig:TimeSeries}J-M). The addition of noise has two biologically interesting effects on the system:

\subsection{Higher secondary amplitude renders the system more robust to noise.}
In order to destroy or contain pathogens, the neutrophils need a distinct concentration gradient to follow to the site of infection. Adding sufficient amount of noise to the period of oscillations and/or in the length of the refractory period of each cell causes the propagating waves to disintegrate or to lose their radial gradient. Biologically, this may compromise the effectiveness of the inflammatory response. 

\begin{figure}[t]
\includegraphics[width=0.5\textwidth]{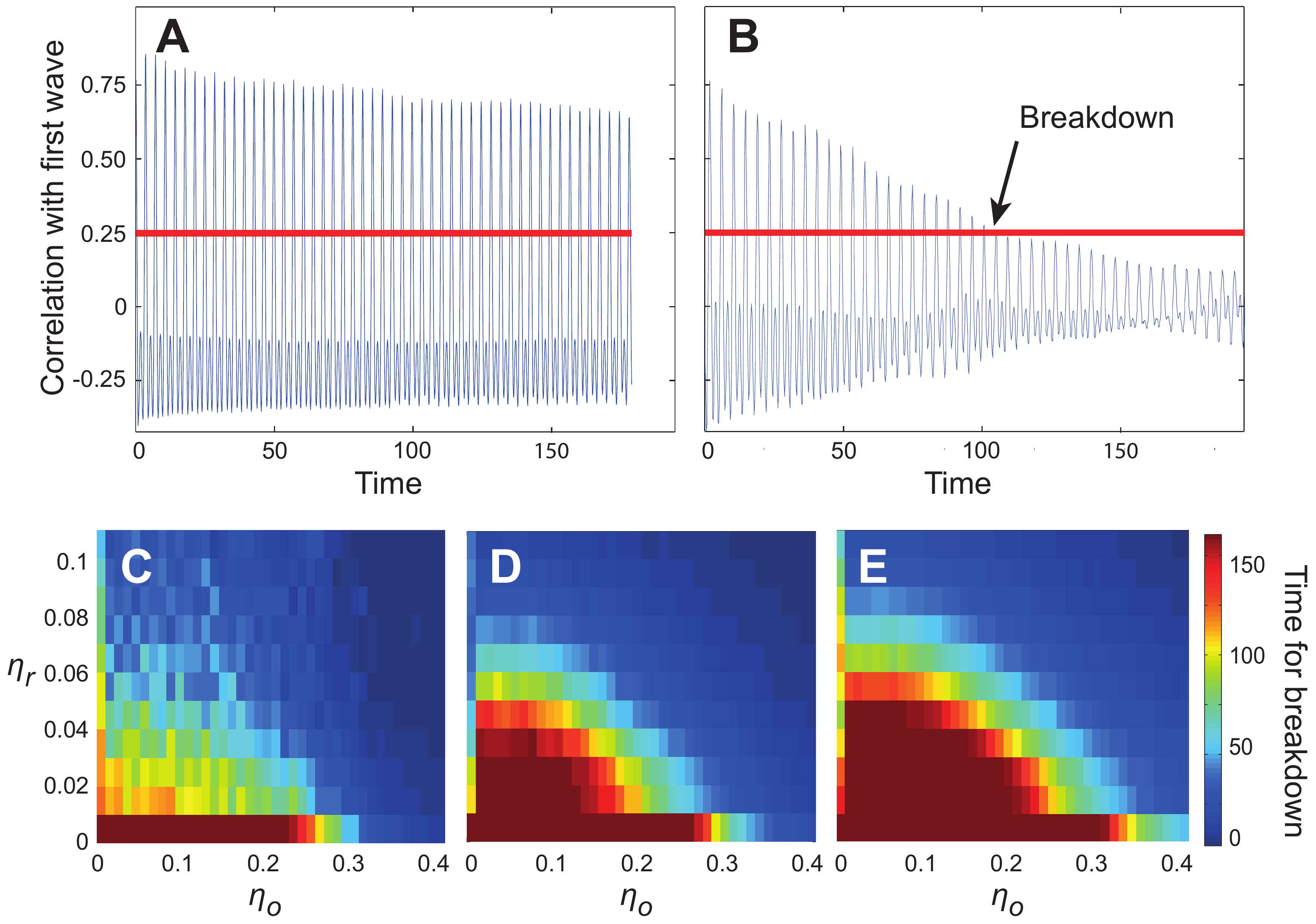}
\caption{\textbf{A-B}: Correlation to the first wave as a function of time in hours, illustrating how the characteristic time for breakdown of wave propagation is found. The red line corresponds to the correlation threshold of 0.25. 
\textbf{A} shows correlations of a system with little noise that is able to sustain waves. Parameters: $A=0.25$, $\eta_o = 0.02$\% and $\eta_r = 0.01$\%.  
\textbf{B} shows correlations of a system with sufficient noise to break the waves. Parameters: $A=0.25$, $\eta_o = 0$\% and $\eta_r = 0.05$\%. 
\textbf{C-E}: Characteristic time for breakdown as function of noise levels in the period of oscillations and in the length of refractory period for different secondary amplitudes. The secondary amplitude in panel C-E is 0.15, 0.25, and 0.35, respectively.}
\label{fig:PS}
\end{figure}

The time it takes the system to lose its wave structure depends on the amount of noise in the system. In order to examine this time before breakdown, the state of the system over time is compared to the state of the system the first time the waves fill the entire grid. This is achieved by calculating the Pearson correlation coefficient between cytokine concentration of the first grid in which the waves have reached the outer edges and the grid in all later time steps: 

\begin{equation}\label{eq:Corr}
C_{1,t} = \langle \frac{g_1- \langle g_1 \rangle}{\sigma_1} \cdot \frac{g_t - \langle g_t \rangle}{\sigma_t} \rangle .
\end{equation}
Here, $g_1$ is the cytokine levels of the first grid that is filled with waves, and $g_t$ is the grid at a later time step. Angle brackets denote spacial average and $\sigma$ denotes spatial standard deviation. The correlation coefficients are plotted as a function of time in Fig.\ref{fig:PS}A-B.
The correlation coefficients decay due to noise destroying the waves. The characteristic time for breakdown of wave structure is defined as the last time the correlation coefficient to the first wave is more than 0.25 (see Fig. \ref{fig:PS}A-B).
The exact value of this threshold is not essential for the main results, as long as it is low enough to allow for some irregularity in wave patterns.

The correlation method allows us to investigate the characteristic time for breakdown as a function of noise in the period the oscillations and noise in the refractory period. 
Interestingly, the secondary amplitude $A$, has a significant effect on the system's ability to sustain wave propagation.
As is seen in Fig. \ref{fig:PS}C-E, the parameter range with stable wave patterns increases with increasing secondary amplitude. Thus, high secondary amplitude makes the system more robust to noise. One possible explanation for this is that secondary waves might help straighten up the lagging sections of the major wavefronts by increasing the cytokine concentration at the inflections points and effectively accelerating them. Since lagging cells can get excited by having \textit{e.g.} two neighbours in primary peak and $1/A$ neighbours in secondary peak, a high secondary amplitude will lead to a larger acceleration. 
There is, however, an upper limit to the secondary amplitude: Waves should only originate from the central stimulus, so the combined secondary amplitudes from the 8 nearest neighbours of a cell should never exceed the activation threshold $\theta = 3$. That is, if the amplitude of secondary oscillations is larger than $\theta/8$, cells have a possibility of getting activated by the secondary oscillations of its neighbours. This could happen anywhere in the tissue, not just close to a stimulus, causing an inflammatory response from healthy tissue leading to chronic inflammation. Although this upper limit on the secondary amplitude is unknown for living cells, it is interesting to note that the experimentally observed secondary amplitude is significantly smaller than the initial peak, with the secondary amplitude being about 20 $\%$ of the first peak \cite{Ashall2009}!

\subsection{Noise leads to self-sustained excitations resembling chronic inflammation.}
To further address the issue of chronic inflammation, we tried removing the stimulus after 120 periods of oscillations, corresponding to 200 hours. If this eliminated the cytokine production, such that no cell in the tissue was re-excited to its first peak after another 48 periods of oscillations, we defined the system to be {\it sensitive} to the removal of the stimulus. If not, the system was declared {\it insensitive}. A system that is insensitive to the removal of the stimulus is not able to relax back to the pre-infected state, and is reminiscent of chronic inflammation. 

\begin{figure}[t!]
 \includegraphics[width=0.5\textwidth]{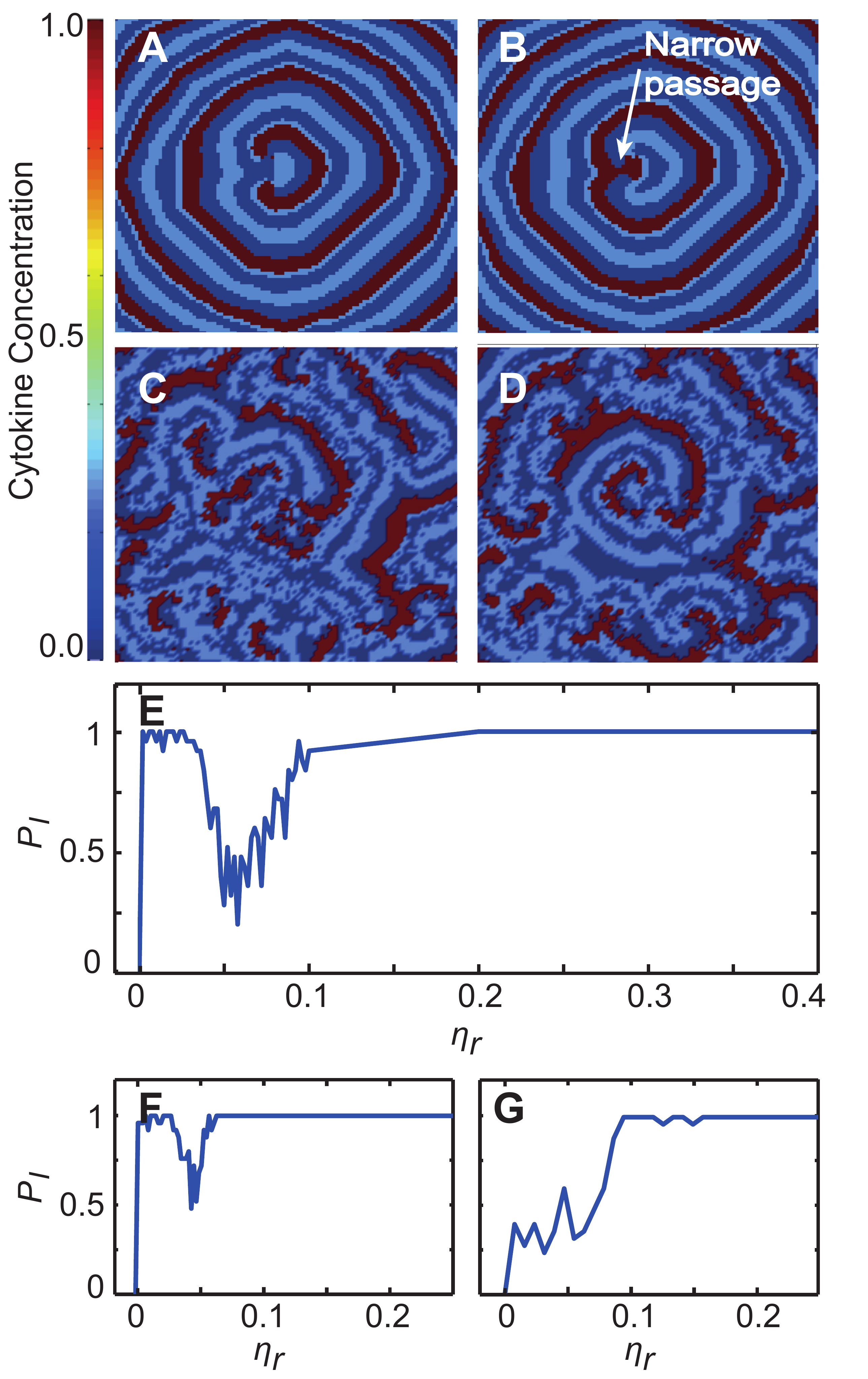}
\caption{{\bf A-D} show two distinct ways the system can become insensitive to the stimulus. {\bf A-B}: For small noise levels, a double spiral appears close to the removed stimulus. To sustain this double spiral, the previous wavefront needs to excite the central cells through a narrow passage. Parameters: $A=0.25$, $\eta_o = 0\%$, $\eta_r=2$\%. {\bf C-D}: For larger noise levels spirals arise numerous places in the tissue. Parameters: $A=0.25$, $\eta_o = 0\%$, $\eta_r=20$\%. {\bf E}: The probability $P(I)$ that the noise in the refractory period causes the system to become insensitive to removal of stimulus as a function of noise in the refractory period. Interestingly, the curve is non-monotonic. {\bf F}: Same as E but with a grid size of $202 \times 202$ cells instead of $101 \times 101$. The larger system size expedites the emergence of stable spirals. {\bf G}: Same as E but with with a noise in the oscillation period of $\eta_o = 10$ \%, which destabilizes the double spirals for low noise levels, making the insensitivity curve monotonic.}
\label{fig:Prob}
\end{figure}

We found that noise in the oscillation period and/or in the duration of refractory period can create stable additional sources and spirals, which excite the system independently of the stimulus, see Fig. \ref{fig:Prob}A-D.
Thus, noise not only destroys the radial gradient of the wave pattern, so that neutrophils will be unable to locate the stimulus, but the system can become insensitive so that removing the stimulus will not terminate the inflammatory response.

As the system is particularly sensitive to noise in the refractory period, we investigate how the insensitivity depends on the level of refractory noise, $\eta_r$. We define the probability of becoming insensitive to the stimulus, $P(I)$, as the fraction out of 30 runs, where the system becomes insensitive.

Interestingly, with no noise in the oscillation period, the probability to become insensitive is a non-monotonous function of refractory noise, as seen in see Fig. \ref{fig:Prob}E. With no noise, the system is sensitive, $P(I)=0$ at $\eta_r=0$, but a small increase in noise has a dramatic effect and completely destroys the sensitivity, $P(I) \approx 1$ when $\eta_r=0.01$. 
At this small noise level, the insensitivity is caused by a double spiral very close to the center, where the stimulus once was, as seen in Fig. \ref{fig:Prob}A-B. To sustain this double spiral, the previous wavefront repeatedly excites the central cells through a narrow passage of cells, as illustrated in Fig. \ref{fig:Prob}B. 
If the noise level is increased further to $0.04<\eta_r<0.1$, this signal can not be transmitted through the narrow passage, and the probability that the system is insensitive decreases. This explains the decline in probability for insensitivity in figure \ref{fig:Prob}E. Increasing noise further, $\eta_r>0.1$ produces spirals in rich numbers everywhere in the tissue, as shown in Fig. \ref{fig:Prob}C-D. These spiral structures are quite stable and can continue to excite cells on long timescales after the stimulus has been removed.

To better understand the non-monotonous behavior of the systems sensitivity, we have tested the case with larger system size, which should give higher chance for spirals to originate away from the center. As expected, Fig. \ref{fig:Prob}F shows that the onset of permanent insensitivity has been expedited to lower noise levels, reducing the range of stimulus sensitivity.
Furthermore, adding noise to the oscillation period prevents the formation of double spirals close to the center, as the cytokine signal is again prevented to pass through the narrow passage of Fig. \ref{fig:Prob}B. As a result, noise in the period of oscillation of $\eta_o = 10$\% turns $P(I)$ into a monotonously increasing function (see Fig. \ref{fig:Prob}G). This also explains why a small noise in the oscillation period drastically extends the characteristic time before breakdown of wave propagation in Fig. \ref{fig:PS}D-E.

\section{Discussion}
The inflammatory response is initiated by the NF-$\kappa$B regulatory network. The dynamics of NF-$\kappa$B response is very peculiar when measured on single cell level. While the oscillatory dynamics of NF-$\kappa$B in single cells have been the focus of extensive research \cite{Ashall2009, Lee2009}, little attention has been given to the the role of the lower secondary amplitude and the noise in oscillations. 

Rather than seeking the single-cell explanations, we have taken the perspective of spatially distributed cells, which collectively propagate cytokine signals through the tissue, thereby recruiting white blood cells from the blood stream to the site of infection. 
This approach led us to several interesting findings:
a) The system is more sensitive to noise in refractory period than in oscillation period.
b) Moderate noise in secondary oscillations stabilizes concentric wave patterns in the presence of noise in the refractory period. The effect is stronger with higher secondary amplitude. However, the secondary amplitude should never be so high that a group of cells in their secondary oscillation can re-excite a cell to its first peak, as this potentially would create numerous sources of propagating waves in the system.
This may help to explain why the experimentally observed secondary amplitude is indeed much smaller than the initial peak, $\sim 20 \%$.
This prediction can be tested experimentally. The propagation of waves could be measured in a 2D culture of mammalian cells with a fluorescent reporter fused to NF-kB \cite{Lee2009}. In this case nuclear NF-kB will serve as a proxy for cytokine production. The secondary amplitude is controlled by A20, such that higher expression of A20 results in lower amplitude. There already exist cell lines where the amounts of A20 can be tuned externally \cite{Werner2005}. Combining the two: fluorescent microscopy of 2D culture of cells with externally tunable A20, one can test our predictions that higher amplitude would stabilize wave propagation. 

c) Increasing noise in refractory period increases the chance of self-sustained excitations, thus decoupling the excitations from the original source. This situation is in effect very similar to chronic inflammation -- inflammation which persists and re-occurs even after the source of damage has been removed \cite{Nathan2010}.
The system can show a very rich, non-monotonous behavior in how it turns insensitive with increasing refractory noise. 
Remarkably, noise in secondary oscillations postpones the onset of insensitivity, thus rendering system more robust to low noise in refractory period.
Our findings related to the effect of noise in the refractory period are limited to the inflammatory response alone, but are general for other excitable media models.

Overall, our investigation shows that both of the experimentally observed features: the lower amplitude of secondary oscillations and the noise in the period of oscillations, are crucial for stabilizing wave patterns as well as maintaining system sensitivity.

%\begin{acknowledgments}
{\it Aknowledgement} This study was supported by the Danish National Research Foundation through the Center for Models of Life and Steno fellowship (AT).
%\end{acknowledgments}

%\bibliographystyle{h-physrev3}
%\bibliographystyle{plain}
%\bibliography{referencesNFkB}

\begin{thebibliography}{10}

\bibitem{Brasier2006}
A.~R. Brasier,
\newblock Cardiovasc Toxicol {\bf 6}, 111 (2006).

\bibitem{Hoffmann2006}
A.~Hoffmann and D.~Baltimore,
\newblock Immunol Rev {\bf 210}, 171 (2006).

\bibitem{Pahl1999}
H.~L. Pahl,
\newblock Oncogene {\bf 18}, 6853 (1999).

\bibitem{Lee2009}
T.~K. Lee {\em et~al.},
\newblock Sci Signal {\bf 2}, ra65 (2009).

\bibitem{Kasper2010}
C.~A. Kasper {\em et~al.},
\newblock Immunity {\bf 33}, 804 (2010).

\bibitem{Yde2011a}
P.~Yde, M.~H. Jensen, and A.~Trusina,
\newblock Phys Rev E Stat Nonlin Soft Matter Phys {\bf 84}, 051913 (2011).

\bibitem{Yde2011c}
P.~Yde, B.~Mengel, M.~H. Jensen, S.~Krishna, and A.~Trusina,
\newblock BMC Syst Biol {\bf 5}, 115 (2011).

\bibitem{Geiger2003}
J.~Geiger, D.~Wessels, and D.~R. Soll,
\newblock Cell Motil Cytoskeleton {\bf 56}, 27 (2003).

\bibitem{Witko-Sarsat2000}
V.~Witko-Sarsat, P.~Rieu, B.~Descamps-Latscha, P.~Lesavre, and
  L.~Halbwachs-Mecarelli,
\newblock Lab Invest {\bf 80}, 617 (2000).

\bibitem{Tay2010}
S.~Tay {\em et~al.},
\newblock Nature {\bf 466}, 267 (2010).

\bibitem{Mengel2012}
B.~Mengel, S.~Krishna, M.~H. Jensen, and A.~Trusina,
\newblock Physica A: Statistical Mechanics and its Applications {\bf 391}, 100
  (2012).

\bibitem{Lipniacki2004}
T.~Lipniacki, P.~Paszek, A.~R. A.~R. Brasier, B.~Luxon, and M.~Kimmel,
\newblock J Theor Biol {\bf 228}, 195 (2004).

\bibitem{Ashall2009}
L.~Ashall {\em et~al.},
\newblock Science {\bf 324}, 242 (2009).

\bibitem{Marcello2010}
M.~Marcello and M.~R.~H. White,
\newblock Biochem Soc Trans {\bf 38}, 1247 (2010).

\bibitem{Turner2010}
D.~A. Turner {\em et~al.},
\newblock J Cell Sci {\bf 123}, 2834 (2010).

\bibitem{Lipniacki2007}
T.~Lipniacki and M.~Kimmel,
\newblock Cardiovasc Toxicol {\bf 7}, 215 (2007).

\bibitem{Wolfram1983}
S.~Wolfram,
\newblock Rev. Mod. Phys. {\bf 55}, 601 (1983).

\bibitem{Bub2002}
G.~Bub, A.~Shrier, and L.~Glass,
\newblock Phys Rev Lett {\bf 88}, 058101 (2002).

\bibitem{Bub2005}
G.~Bub, A.~Shrier, and L.~Glass,
\newblock Phys Rev Lett {\bf 94}, 028105 (2005).

\bibitem{Sawai2005}
S.~Sawai, P.~A. Thomason, and E.~C. Cox,
\newblock Nature {\bf 433}, 323 (2005).

\bibitem{Marr2006}
C.~Marr and M.-T. Hutt,
\newblock Physics Letters A {\bf 349}, 302 (2006).

\bibitem{Nelson2004a}
D.~E. Nelson {\em et~al.},
\newblock Science {\bf 306}, 704 (2004).

\bibitem{Werner2005}
S.~L. Werner, D.~Barken, and A.~Hoffmann,
\newblock Science {\bf 309}, 1857 (2005).

\bibitem{Nathan2010}
C.~Nathan and A.~Ding,
\newblock Cell {\bf 140}, 871 (2010).

\end{thebibliography}

\end{document}